\documentclass[conference]{IEEEtran}
\usepackage{cite}
\usepackage[T1]{fontenc}
\usepackage{threeparttable}
\usepackage[latin9]{inputenc}
\usepackage{amsmath}
\PassOptionsToPackage{draft}{hyperref}
\usepackage[bookmarks=false]{hyperref}
\PassOptionsToPackage{bookmarks=false}{hyperref}
\usepackage[output-decimal-marker={.}]{siunitx}
\PassOptionsToPackage{bookmarks=false}{hyperref}
\usepackage{array}
\PassOptionsToPackage{draft}{hyperref}
\usepackage[left=0.64in,right=0.64in,top=0.64in,bottom=0.64in]{geometry}

\newcolumntype{C}{>{$}c<{$}}
\usepackage{booktabs}
\usepackage{caption}
\usepackage{graphics}
\newsavebox{\foobox}

\usepackage{bm}
\usepackage{makecell}
\usepackage{tikz}
\usetikzlibrary{shapes.arrows}
\usepackage{color}
\usepackage{xcolor}
\usepackage{array}
\usepackage{booktabs}
\usepackage{textcomp}
\usepackage{multirow}
\usepackage{mathtools}

\usepackage{url}
\usepackage{pifont}
\usepackage{amsfonts}
\usepackage{indentfirst}
\usepackage{amsthm}
\usepackage{amssymb}
\usepackage[export]{adjustbox}
\usepackage{verbatim}
\usepackage{graphicx}
\usepackage{mathrsfs} 
\DeclareMathAlphabet{\mathpzc}{OT1}{pzc}{m}{it}
\usepackage{booktabs}

\usepackage{color}

\usepackage[ruled,linesnumbered]{algorithm2e}

\definecolor{darkgreen}{rgb}{0, 0.5, 0}
\SetAlFnt{\small\color{black}\tt}
\LinesNumbered

\usepackage{graphicx}
\usepackage{placeins}

\newcounter{defcounter}
\usepackage[shortlabels]{enumitem}
\usepackage{cite}
\usepackage[ruled]{algorithm2e}
\mathchardef\period=\mathcode`.
\DeclareMathSymbol{.}{\mathord}{letters}{"3B}
\usepackage{tikz}
\tikzstyle{io} = [fill=black,inner sep=2pt,circle]

\graphicspath{ {images/} }

\usetikzlibrary{arrows,positioning}
\usetikzlibrary{shapes.geometric,positioning,shapes,arrows,calc, decorations.pathreplacing, arrows.meta}

\everymath{\displaystyle}
\usetikzlibrary{backgrounds}

\usepackage{pifont}

\makeatletter
\def\endthebibliography{%
	\def\@noitemerr{\@latex@warning{Empty `thebibliography' environment}}%
	\endlist
}

\makeatletter
\newcommand*\bigcdot{\mathpalette\bigcdot@{.5}}
\newcommand*\bigcdot@[2]{\mathbin{\vcenter{\hbox{\scalebox{#2}{$\m@th#1\bullet$}}}}}
\makeatother

\usepackage{xparse,xcolor}

\ExplSyntaxOn

\NewDocumentCommand{\setupbibcolors}{m}
 {
  \cs_set_protected:Npn \bibitem ##1
   {
    \color{ \str_case:nnF { ##1 } { #1 } { black } }
    \heba_bibitem:n { ##1 }
   }
 }

\cs_set_eq:NN \heba_bibitem:n \bibitem

\ExplSyntaxOff

\setupbibcolors{
  {baza3}{white}
  {baza1}{white}
  {baza2}{white}
  {baza4}{white}
   {baza5}{white}
  {baza6}{white}
  {baza7}{white}
  {baza8}{white}
   {baza9}{white}
    {baza10}{white}
    {baza11}{white}
    {baza12}{white}
    {baza13}{white}
}

\makeatletter

\theoremstyle{plain}

\usepackage{cite} 
\usetikzlibrary{shapes,arrows}
\tikzstyle{line}=[draw] 
\usepackage{algorithmic}
\usepackage{graphicx}
\usepackage{enumitem}

\usepackage{url}
\usepackage{soul}
\usepackage{xcolor}

\makeatother
\usepackage[english]{babel}
\providecommand{\theoremname}{Theorem}

\begin{document}

\title {Mimic Learning to Generate a Shareable Network Intrusion Detection Model}

\author{\IEEEauthorblockN{
  Ahmed Shafee\IEEEauthorrefmark{1}$^1$,
  Mohamed Baza\IEEEauthorrefmark{1}$^2$,
  Douglas A. Talbert\IEEEauthorrefmark{2}$^3$,
  Mostafa M. Fouda\IEEEauthorrefmark{1}\IEEEauthorrefmark{4}$^4$,
  Mahmoud Nabil\IEEEauthorrefmark{3}$^5$, and \\
  Mohamed Mahmoud\IEEEauthorrefmark{1}$^6$}
  
  \IEEEauthorblockA{%
    \IEEEauthorrefmark{1}Department of Electrical and Computer Engineering, Tennessee Tech University, Cookeville, TN, USA
  }
   \IEEEauthorblockA{%
    \IEEEauthorrefmark{2}Department of Computer Science, Tennessee Tech University, Cookeville, TN, USA
  }
   \IEEEauthorblockA{%
    \IEEEauthorrefmark{4}Department of Electrical Engineering, Faculty of Engineering at Shoubra, Benha University, Egypt
  }
    \IEEEauthorblockA{%
    \IEEEauthorrefmark{3}Department of Electrical and Computer Engineering, North Carolina A\&T State University, Greensboro, NC, USA
  }
  
			Emails:  \{aashafee42$^1$, mibaza42$^2$\}@students.tntech.edu, \{dtalbert$^3$, mfouda$^4$, mmahmoud$^6$\}@tntech.edu,\\mnmahmoud@ncat.edu$^5$
}

\maketitle
\IEEEpeerreviewmaketitle

 \begin{abstract}
Purveyors of malicious network attacks continue to increase the complexity and the sophistication of their techniques, and their ability to evade detection continues to improve as well. Hence, intrusion detection systems must also evolve to meet these increasingly challenging threats. Machine learning is often used to support this needed improvement. However, training a good pwhiteiction model can require a large set of labeled training data. Such datasets are difficult to obtain because privacy concerns prevent the majority of intrusion detection agencies from sharing their sensitive data. In this paper, we propose the use of mimic learning to enable the transfer of intrusion detection knowledge through a teacher model trained on private data to a student model. This student model provides a mean of publicly sharing knowledge extracted from private data without sharing the data itself. Our results confirm that the proposed scheme can produce a student intrusion detection model that mimics the teacher model without requiring access to the original dataset.

	\end{abstract}

	\begin{IEEEkeywords}

Machine learning, Mimic learning, Intrusion detection system

	\end{IEEEkeywords}
	\vspace{-0.1in}
\section{Introduction}
\subsection{Motivation}

The Internet has become an essential tool in our daily lives. It aids people in many areas, such as business, entertainment, and education~\cite{Groan2005}. Along with such benefits, however, comes the ever-present risks of network attacks. Thus, many systems have been designed to block such attacks. One such danger is that of malicious software (malware) that can be used by hackers to compromise a victim's machine ~\cite{Stallings:2008:OSI:1816910}. Millions of dollars are lost each year because of Ransomware, botnets, backdoors and Trojans malware. Therefore, network security is a serious concern, and intrusion detection is a significant research problem impacting both business and personal networks.

This work focuses on intrusion detection systems (IDSs). IDSs assist network administrators in detecting and preventing external attacks. That is, the goal of an IDS is to provide a wall of defense that stops the attacks of online intruders. IDSs can be used to detect different types of malicious network communications and computer system usage, better than a conventional firewall. Of particular interest to us are anomaly intrusion detection systems. Such systems are based on the assumption that the behavior of intruders differs from that of a legitimate user.

One effective technique in anomaly IDS is by using machine learning to detect unusual patterns that could suggest that an attack is happening. The ultimate goal of these techniques is to determine whether deviation from the established normal usage patterns can be flagged as intrusions \cite{Garcia:2014:SNB:2905797.2905803}. In the literature, some works apply single learning techniques, such as neural networks \cite{article2}, genetic algorithms, or support vector machines. Other systems are based on a combination of different learning techniques, such as hybrid or ensemble techniques. 

\vspace{4 pt}

While machine learning has been widely adopted by large intrusion detection agencies (IDAs) such as Kaspersky \cite{MICyber}, McAfee, and Norton, some challenges have not yet been fully addressed. Learning algorithms benefit from large training sets. However, finding large datasets that contain \textit{well defined malware with the latest up-to-date signature and zero-day attacks} is not an easy task, and typically, it requires a specialized IDA, e.g., Kaspersky
or
McAfee.
The lack of such data in large quantities can result in an inaccurate detection model because of insufficient training.  

Ideally, such agencies could share their data with researchers working to develop improved detection systems. However, the private nature of this data prevents it from being shared. One na\"{i}ve solution is to allow these agencies to share their prediction models. However, such models are known to implicitly memorize details from the training data and, thus, inadvertently reveal those details during inference~\cite{DBLP2}. Moreover, some organizations might use the model to infer sensitive information about new malware that has been detected. What is more, that inferred information might be misused to create new malware. \textit{Such risks prevent the IDAs from sharing their models because of their justifiable concerns regarding keeping their data private}. 


Recently, the idea of \textit{mimic learning} has been introduced as a solution for preserving the privacy of sensitive data\cite{DBLP2}. Mimic learning involves labeling previously unlabeled public data using models that are trained on the original sensitive data. After that, a new model is trained on the newly labeled public data to produce a prediction model that can be shared without revealing the sensitive data itself or exposing the model that is directly trained on this data
. Therefore, it is suggested, mimic learning can enable the transfer of knowledge from sensitive and private data to a shareable model without putting the data's privacy at risk.


\subsection{Related Work} Several schemes have been proposed to study how machine learning can be used for intrusion detection systems\cite{6997492}. However, there has been little research on the use of mimic learning to enable sharing trained models instead of the original sensitive data as an option for transferring knowledge
~\cite{Papernot2017SemisupervisedKT}. One example of mimic learning research is in the area of \textit{information retrieval} (IR), which seeks to effectively identify which information resources in a collection are relevant to a specific need or query. In IR applications, having access to large-scale datasets is essential for designing effective systems. However, due to the sensitivity of data and privacy issues, not all researchers have access to such large-scale datasets for training their models. Motivated by previous challenges, Dehghani, et al.~\cite{DBLP2} proposed a mimic learning scheme to train a 
shareable model using weak- and full-supervision techniques~\cite{dehghani2017neural} on the data. Then, this trained model can safely transfer knowledge by enabling a large set of unlabeled data to be labeled, thereby, creating the needed large-scale datasets without having to share sensitive, private data. Unfortunately, current research work has not yet studied the use of mimic learning in intrusion detection system which comprises a domain specific environment including various attributes and requires new prediction models.

\subsection{Main Contribution}
In this paper, we propose a mimic learning approach that enables intrusion detection agencies to generate shareable models that facilitate knowledge transfer among organizations and research communities. \textit{To the best of our knowledge, this work is the first to examine the use of mimic learning in the area of intrusion detection.}

Thus, the main contribution of this paper is \textbf{the introduction and empirical evaluation of a mimic learning scheme for intrusion detection datasets that effectively transfers knowledge from 
a sensitive unshareable dataset to a shareable predictive model}. 






The remainder of this paper is organized as follows. Section~\ref{System Model} describes the system model. The background is presented in \ref{Background}. Our proposed scheme is discussed in details in Section~\ref{Proposed scheme}. The experimental results are discussed in Section~\ref{exp}.  Our conclusions are presented in Section~\ref{Conclusion}.

\section{System Model}
\label{System Model}

In this section, we present the considered system model followed by intrusion detection system (IDS) framework.
\label{System Model and Preliminaries}

	\begin{figure}[!t]
  \begin{center}
   \includegraphics[width=\columnwidth]{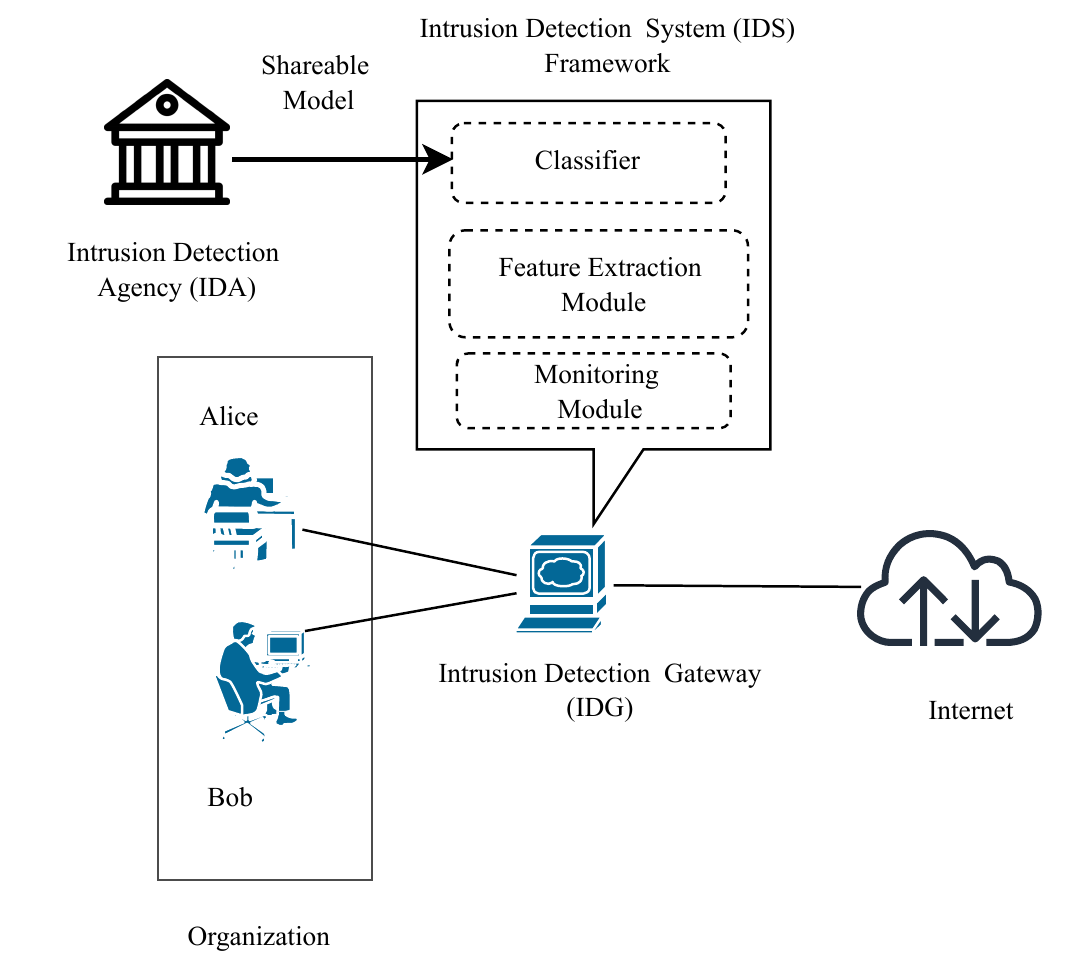}
        \caption{Illustration of the system model under consideration, where there is an organization with some users and IDG. Also, there is IDA which provides the shareable model.}
\label{fig: netmodel}
\end{center}
\end{figure}

\subsection{System Model}
 As depicted in Fig.~\ref{fig: netmodel}, the assumed system model has the following entities.
   
    \begin{itemize}
       \item \textit{Intrusion Detection Agencies (IDAs)}: IDAs are agencies such as cybersecurity and anti-virus providers. They are responsible for developing  antivirus, Internet security, endpoint security, and other cybersecurity products and services. Examples of IDAs include Kaspersky Labs, McAfee, and Norton. 
        
        \item \textit{Organizations}: Companies, universities, and government agencies that are connected to the Internet and have the potential to encounter daily cyber attacks.
        
        \item \textit{Intrusion Detection Gateway (IDG)}: IDGs are independent entities within each organization that are responsible for detecting malicious software attacks. An IDG manages the intrusion detection system (IDS) framework.

\end{itemize}
\subsection{Intrusion Detection System Framework}
An intrusion detection system is a system that is responsible for monitoring, capturing, and analyzing events occurring in computer systems and networks to detect intrusion signals\cite{Liao2013IntrusionDS}. 
In this paper, we consider the following IDS framework that consists of three modules as shown in Fig.~\ref{fig: netmodel}\cite{NetworkRATDetection}.

 \begin{enumerate}
 
\item \textit{Monitoring Module}: This module is responsible for monitoring network packets that pass through the network gateway. 

\item \textit{Feature Extraction Module}: This module is responsible for extracting the features that describe each feature vector. 
In this paper, a set of 
statistical features (e.g. number of source bytes, number of destination bytes, and protocol) were extracted for each 
feature vector to represent the behavior of the network traffic flows.
Statistical features are chosen to avoid privacy concerns among users as well as problems associated with encrypted data.

\item \textit{Classifier Module.} This module is responsible for classifying a given feature vector as either benign or malicious based on the extracted features for that vector. Supervised machine learning algorithms are used in our scheme for the task of classification.
 \end{enumerate}
The monitoring and feature extraction modules are the same as in \cite{NetworkRATDetection}, however, in this paper we focus on the classifier module.

\section{Background}
\label{Background}
In this section, we present an introduction to supervised machine learning techniques that are needed to understand the way the classifier module works.

Supervised machine learning algorithms take, as input, data labeled with either a numeric or categorical value and produce a program or model capable of using patterns present in the input data to guide the labeling of new or previously unseen data. In this paper, our classifiers use a categorical label with a value of either \textit{benign} or \textit{malicious}.

Our experiments include the following four machine learning
algorithms for classifying connections as benign or malicious: Decision Tree Induction (DT), Random Forests (RF), Support Vector Machine (SVM), and Na\"{i}ve Bayes (NB). We provide an overview of each of these algorithms.

\begin{itemize}

\item \textit{Decision Tree Induction (DT)}:
A decision tree organizes a hierarchical sequence of questions that lead to a classification decision. Decision tree induction creates this tree through a recursive partitioning approach that, at each step, selects the feature whose values it finds most useful for predicting labels at that level in the hierarchical structure, partitions the data according to the associated values, and repeat this process on each of the resulting nodes until a stopping condition is met. Once built, new data is classified by using its feature values to guide  tree traversal from the root to a leaf, where a class label is assigned. At each node, a given feature from the sample is evaluated to decide which branch is taken along its path to a leaf. Different algorithms use different criteria for selecting the feature at each node in the tree. As an example, we look at ID3~\cite{Quinlan1986}.

To build the tree, ID3 uses a greedy method that relies on entropy and information gain (IG)~\cite{Kruegel2003}
to determine the best feature at each node as given in Eq. \ref{eq:1-2}. 
\begin{equation}
IG(D,A)=H(D)-\sum_{t\text{\ensuremath{\in}T}}p(t)H(t),\label{eq:1-2}
\end{equation}

where

\begin{equation}
H(D)=-\sum_{y\text{\ensuremath{\in}T}}p(y)log_{2}p(y),\label{eq:2-1}
\end{equation}

where $D$ is the current data set, $t$ is the subset produced from
$D$ after splitting it based on attribute $A$, $y$ is the set of classes,
$IG(A,D)$ is the information gain of the system at attribute $A$, $H(D)$ is the system entropy, $H(t)$ is the entropy of each subset generated as a result of splitting the set $D$ using attribute
$A$, $T$ is the set that contains all the subsets that are generated from partitioning set $D$ by attribute $A$, and $p(t)$ is the ratio of the number of elements in $t$ to the number of elements in $D$.
    
\item \textit{Random Forests (RF)}: To increase the accuracy and stability of decision trees, RF leverages a bagging technique \cite{Breiman1996} to generate a collection (or forest) of trees. The label is determined either by averaging the output decisions in the case of numerical labels or, in the case of categorical labels, by the class with the maximum number of ``votes'' (the class selected by a majority of the tree in the forest)~\cite{Breiman2001}. 

\item \textit{Support Vector Machine (SVM):} A SVM \cite{Cortes:1995:SN:218919.218929} is a machine
learning algorithm that attempts to separate points of data in $n$-dimensional space using a hyperplane of $n-1$ dimensions. The hyperplane provides the best separation when the distance of the nearest points to the plane is maximal for both sides \cite{Shafiq2009}. We include SVM because of its scalability and high accuracy in complex classification problems \cite{article}.

\item \textit{Naive Bayes (NB).} It is a statistical classification methodology established on Bayes theorem\cite{6156687}. It is based on the assumption that the features of a given sample are conditionally independent from each other. This assumption enables a tractable calculation of the posterior probability of class $C$ for a data
sample with $n$ attributes. Given a data $X$ with 
features $X=\{x_{1},\ldots,x_{i},\ldots,x_{v}\}$ and a set of classifiers$\{C_1,\ldots,C_K\}$, NB algorithm identifies the class with the maximum posterior probability distribution 
$y$ as given in Eq.~\ref{eq:3-1}.


\begin{equation}
    \begin{aligned}
        z = \underset{k\epsilon\{1,\ldots,K\}}{\textit{argmax }} P(C_{k}|X)\label{eq:3-1}
    \end{aligned}
\end{equation}

where
\begin{equation}
P(C_{k}|X)=P(C_{k})\prod_{i=1}^{n}P(x_{i}|C_{k})\label{eq:3-2}
\end{equation}

where $z$ is the label assigned by the NB classifier, $P(C_k|X)$ is the posterior probability of class $C_k$, $P(x_i|C_k)$ is the likelihood probability of feature $x_i$ given class $C_k$, and $P(C_k)$ is the prior probability of class $C_k$. 

\end{itemize}

\section{Proposed Scheme}

\label{Proposed scheme}
In this section, we present the proposed scheme that enables us to generate a shareable classifier model that could be shared with the research communities without sharing the original dataset.
First, a classifier is trained on the original sensitive data to produce a \textit{teacher model}. Then, the generated teacher model is used to annotate publicly available unlabeled data and convert it into labeled data. Next, a new classifier is trained on this newly labeled data to produce a \textit{student model}. 
The proposed methodology is illustrated in Algorithm ~\ref{alg:contract}. In the following subsections, a more detailed explanation is given for generating the teacher and student models.


\begin{algorithm*}[!h]
\SetKwProg{Pn}{function}{:}{end}
\SetKwProg{Fn}{function}{}{}
\SetKwProg{Function}{Function}{}{}
\SetKwData{NumOfUpdatedObjects}{numOfUpdatedObjects}
\SetKwIF{If}{ElseIf}{Else}{if}{}{else if}{else}{end if}
\SetKwFunction{TeacherModelGeneration}{TeacherModelGeneration}
\SetKwFunction{StudentModelGeneration}{StudentModelGeneration}
\SetKwFunction{KnowledgeTransfer}{KnowledgeTransfer}
\SetKwFunction{Refund}{Refund}

\Pn{\TeacherModelGeneration{Sensitive\_Data}}{
      $C \leftarrow  \{classifier\_1$, \ldots ,  $classifier\_m\}$  
	 
Best\_Classifier $\leftarrow$  SelectBest($C$, Sensitive\_Data) \\ 
    Teacher\_Model $\leftarrow$  TrainClassifier(Best\_Classifier, Sensitive\_Data) \\ 

return Teacher\_Model
  \BlankLine
  }

\Pn{\StudentModelGeneration{Public\_labeled\_Data}}{
     $C \leftarrow  \{classifier\_1$, \ldots ,  $classifier\_n\}$  
	 
Best\_Classifier $\leftarrow$  SelectBest($C$, Public\_labeled\_Data) \\
    Student\_Model $\leftarrow$  TrainClassifier(Best\_Classifier, Public\_labeled\_Data)  \\ 
return Student\_Model
  \BlankLine
  }
  \Pn{Main}{
   Sensitive\_Data = GenerateData()  \\
    
TeacherModel = \TeacherModelGeneration(Sensitive\_Data)

Public\_labeled\_Data = AnnotateData(TeacherModel, UnlabeledPublicData) \\
StudentModel = StudentModelGeneration(Public\_labeled\_Data)  \\
relative\_Score\_Difference = EvaluateModels(TeacherModel,StudentModel,Test\_Data) 

\If{ relative\_Score\_Difference < thresholds }{
     
NetworkIntrusionDetection(StudentModel)
  
  }
  }

\caption{Pseudocode for the \textit{Mimic learning} Algorithm}\label{alg:contract}
\label{alg2}
\end{algorithm*}


\subsection{Teacher Model Generation}

\begin{figure}[!t]
  	 \centering
	\captionsetup{justification=centering}
   \includegraphics[width=0.9\columnwidth]{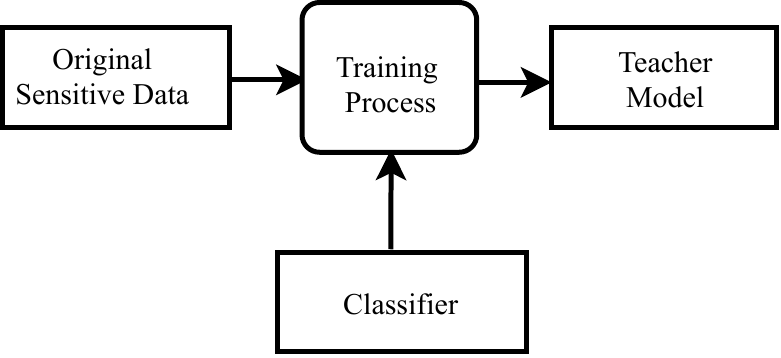}
        \caption{Teacher model training process.}
\label{fig: teacher}
\end{figure}

The process of teacher model generation is shown in Fig. \ref{fig: teacher} where the IDA uses its original sensitive data with a set of classifiers to generate multiple models and the most accurate model is selected as the teacher model.

As depicted in Algorithm \ref{alg:contract}, the process of generating the teacher model starts by evaluating several classifiers to select the classifier with the best performance (see lines 2-3 in Algorithm~\ref{alg:contract}). 
The classifier with the best performance is selected to be the one that is trained on the original sensitive data to produce the teacher model (see lines 4-5 in Algorithm~\ref{alg:contract}). 

\begin{figure}[!t]
  \centering
  \captionsetup{justification=centering}
    \includegraphics[width=1\columnwidth]{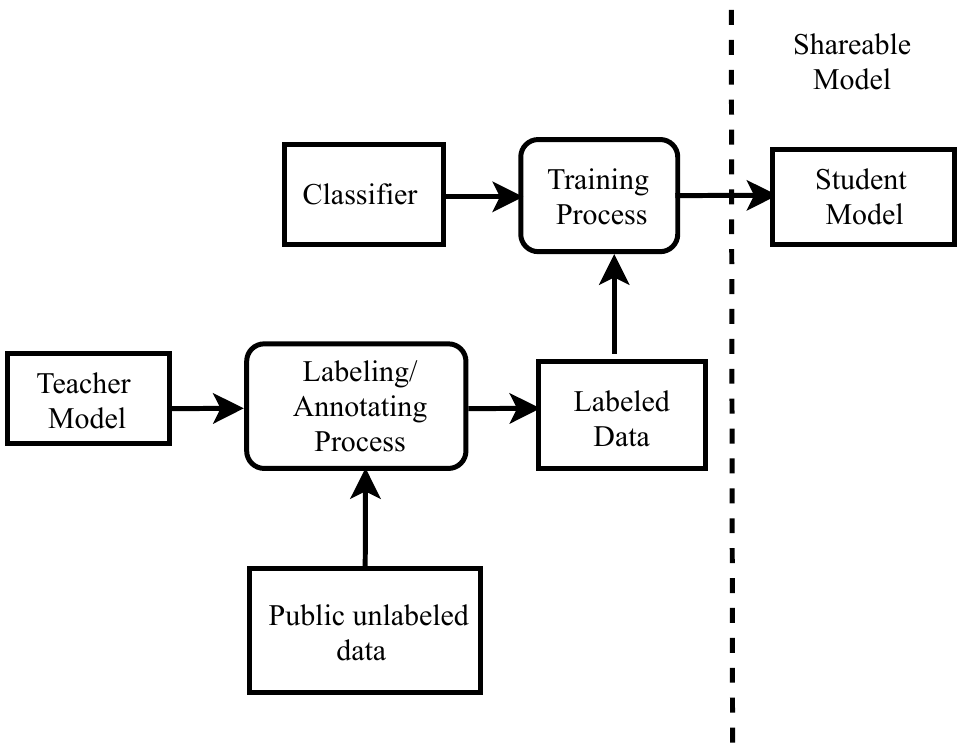}
        \caption{Student model training process.}
\label{fig: student}
\end{figure}

\subsection{Student Model Generation}

As illustrated in Fig. \ref{fig: student}, the process of generating the \textit{student model} starts by using the teacher model to label (or annotate) an unlabeled public dataset to produce newly labeled training data. 
After using the teacher model to label the unlabeled data, a classifier selection and training process, similar to that used with the teacher model, is used to produce our student model (see lines 7-12 in Algorithm \ref{alg:contract}).
This student model training process can be considered a knowledge transfer process. The goal is for the knowledge that the teacher model extracted from the sensitive, private data to be passed along to the student model through the publicly available data that the teacher model labeled. This is illustrated in lines 14-17 in  Algorithm \ref{alg:contract}. Upon training the student model, we propose to evaluate its performance against the teacher model to verify its accuracy before sharing it as shown in lines 18-21 in Algorithm \ref{alg:contract}.

\section{Experiments}
\label{exp}

In this section, we explain the data and the process used to evaluate our scheme. We also discuss the results of this evaluation.


\subsection{Experimental Data}
A total of 136,000 feature vectors were obtained from Kaggle's Open Datasets \cite{Kaggle}, for the purpose of training and evaluating our scheme. This dataset is chosen because it contains various network attacks in addition to a large number of normal traces that mimic the behavior of a normal user, which enables the creation of a more accurate and robust detection model. Each feature vector is described using 41 features.
The dataset contains a combination of benign and malicious data traffic. The benign data resembles the behavior of a normal user inside the monitored network. The malicious data includes the behavior of traffic flows during different kinds of intrusion attacks, e.g., Denial of Service (DoS), Probe, User to Root (U2R), and Remote to User (R2L) attacks.

A brief explanation of each of these attacks is as follows:
\begin{itemize}
\item\textbf{Denial of Service (DoS) attack}: It is a cybersecurity attack in which the intruder tries to prevent the normal usage of the network resources (machines)~\cite{Stallings}. In other words, it tries to exhaust the resource to prevent them from being able to serve the requests of legitimate users.
\item\textbf{Probe attack}: In this type of attack, the intruder tries to scan the network computers to search for any vulnerability to exploit it to compromise the system~\cite{Paliwal2012}.
\item\textbf{User to Root (U2R) attack}: The intruder in this attack tries to exploit the vulnerabilities of a given machine to gain root access~\cite{Paliwal2012}.
\item\textbf{Remote to User (R2L) attack}: It is a cybersecurity attack in which the attacker tries to gain access to a machine that he/she does not have access to~\cite{Paliwal2012}.
\end{itemize}

For the original Kaggle competition, the dataset had a specific test set. The labels for this test set have  been released, and for our experiments, we combined all the data from Kaggle into a single dataset from which different random data sets could be generated as needed. The obtained dataset was divided into three parts. The first part is used as the labeled dataset upon which the teacher model is trained, and it consists of 57,900 feature vectors. The second part was used as the unlabeled dataset by deleting the label column from it, and it consists of 57,900 unlabeled feature vectors. The third part is the test set. It consists of 20,173 feature vectors and constitutes the dataset upon which both the teacher and the student models are evaluated.

\subsection{Experiment Methodology}
The training process of the teacher and student models pass through several steps that can be summarized as follows:
\begin{itemize}
\item \textbf{Step 1: Training the teacher model.} The teacher model is trained offline using a 57,900 feature vector for different network traffic flows representing benign and malicious data (the labeled data). The classifier set is assumed to contain the four classifiers mentioned in section \ref{System Model and Preliminaries}. 
To evaluate the performance of each classifier, we use 10-fold Cross-Validation (CV). $k$-fold CV is the process of dividing the training data into $k$ equal folds (parts). After that, the model is trained on the $k-1$ folds and evaluated on the remaining fold. This operation is repeated $k$ times with each fold being used once as the test data. The $k$ parameter in our experiments is selected to be 10 \cite{Han:2011:DMC:1972541}. 
The classifier with the best performance is 
used in the training process of the teacher model which is then used to label the unlabeled data. 
\item \textbf{Step 2: Labeling/Annotation process.} The unlabeled data is annotated by the teacher model generated in Step 1. The output of this step is a labeled dataset (57,900 feature vectors).
\item \textbf{{Step 3: Training the student model.}}  In this step, the annotated data from Step 2 is used to select and train the student model. The same four classifiers are evaluated at this step using 10-fold CV. The classifier with the best performance is selected in the training process of the student model that should be shared.
\item \textbf{{Step 4: Teacher/Student model evaluation.}} In this step, both the teacher and student models are evaluated using the same test data to compare their performance.
\end{itemize}

\subsection{Considered Key Performance Metrics}

We define the following key performance metrics used in our evaluation process: 
\begin{itemize}
\item \textsl{Detection Accuracy (ACC): } The ratio of the number of true positives and true negatives over the whole number of samples.
\begin{equation}
\textsl{Accuracy (ACC)}=\frac{TP+TN}{TP+TN+FP+FN},\label{eq:1-1}
\end{equation}

where $TP$ represents true positives (the number of malicious samples that are correctly classified as malicious), $TN$ is the true negatives (the number of benign samples that are correctly classified as benign), $FP$ denotes false positives (the number of benign samples incorrectly classified as malicious), and $FN$ represents false negatives (the number of malicious samples incorrectly classified as benign).
\\[2pt]

\item \textsl{True Positive Rate (TPR):} The ratio of the true positives to the total number of malicious samples 
\begin{equation}
\textsl{True Positive Rate (TPR)}=\frac{TP}{TP+FN}\label{eq:2}
\end{equation}

\item \textsl{False Positive Rate (FPR):} The ratio of the false positives to the total number of benign samples 
\begin{equation}
\textsl{False Positive Rate (FPR)}=\frac{FP}{FP+TN}\label{eq:3}
\end{equation}

\item \textsl{True Negative Rate (TNR):} The ratio of the true negatives to the total number of benign samples 
\begin{equation}
\textsl{True Negative Rate (TNR)}=\frac{TN}{TN+FP}\label{eq:1}
\end{equation}

\item \textsl{False Negative Rate (FNR):} The ratio of the false negatives to the total number of malicious samples 

\begin{equation}
\textsl{False Negative Rate (FNR)}=\frac{FN}{FN+TP}\label{eq:5}
\end{equation}
\item \textsl{Area Under Curve (AUC):} It is the area under the receiver operating curve (ROC). The ROC curve is a graphical way that captures the relationship between TPR and FPR \cite{NetworkRATDetection}.  
\end{itemize}
\subsection{Results and Discussion}

The results of evaluating the four classifiers using 10-fold CV for selecting the teacher model are given in Table ~\ref{tab:Performance-Evaluation-of_teacher}. We observe that the RF classifier performs better than other classifiers when it is evaluated on the original sensitive data and that the NB classifier provides the lowest accuracy.  
Thus, the RF classifier is selected for the process of training the teacher model.

\begin{table}[!t]
		\caption{Experimental results of the teacher model.}
		\label{tab:Performance-Evaluation-of_teacher}
		\centering
\scalebox{0.86}{
\begin{tabular}{l|l|l|l|l|l|l}
				
\hline \hline
Classifier & ACC (\%) & TPR (\%) & FPR (\%) & TNR (\%) & FNR (\%) & AUC \\\hline
DT & 99.49 & 99.40 & 0.60 & 99.60 & 0.40 & 0.99 \\\hline
\textbf{RF} & \textbf{99.64}  & \textbf{99.50} & \textbf{0.50} & \textbf{99.80} & \textbf{0.20} & \textbf{1.00} \\\hline
SVM & 94.26 & 91.35 & 8.65 & 96.75 & 3.25 & 0.92 \\\hline
NB & 88.10 & 84.15 & 15.85 & 91.60 & 8.40 & 0.94 \\\hline

\end{tabular} 

}
	\end{table}

The 10-fold CV student model results 
on the annotated public data 
are given in Table \ref{tab:Performance-Evaluation-of_student}. Similarly, the RF classifier performs the best among this set of classifiers. Thus, it is chosen in the training process of the student model. 

The results of the student model (RF classifier) could cause confusion at the first glance as it indicates a better performance for the student model over the teacher model, however both results could not be compared as both models (the teacher and the student) are trained and evaluated on two different datasets. The teacher model is evaluated on the original sensitive data and the student model is developed on the public data that has been labeled by the teacher. For evaluation purposes, a separate test set is used as mentioned in \textbf{step 4}.

The comparison of both the teacher and the student models on the test data is given in Table \ref{tab:Performance-Evaluation-of}. 
The performance of both the teacher and the student models is nearly identical, which supports our assertion that unlabeled data trained by a teacher model can be used to transfer knowledge to a student model without revealing data that is considered sensitive.
It also suggests that, by sharing student prediction models, intrusion detection agencies could enable research communities to benefit from their datasets without directly sharing that sensitive data.
To the best of our knowledge, our proposed schema is the first to consider applying the mimic learning methodology in the intrusion detection field and prove its applicability. Our work differs from what is introduced by Dehghani, et al. \cite{DBLP2} which considers creating a shareable neural ranker mode in the IR application. As a result, the comparison between our proposal and their work is infeasible due to the use of different datasets and performance metrics.\footnote{~\cite{baza1,baza7,baza2,baza8,baza3,baza9,baza4,baza6,baza11,baza13,baza5,baza10}} 


\begin{table}[!t]
		\caption{Experimental results of the student model.}
		\label{tab:Performance-Evaluation-of_student}
		\centering
	\scalebox{0.86}{	
\begin{tabular}{l|l|l|l|l|l|l}
				
\hline \hline
Classifier & ACC (\%) & TPR (\%) & FPR (\%) & TNR (\%) & FNR (\%) & AUC \\\hline
DT & 99.63 & 99.60 & 0.40 & 99.65 & 0.35 & 0.994 \\\hline
\textbf{RF} & \textbf{99.83}  & \textbf{99.8} & \textbf{0.20} & \textbf{99.90} & \textbf{0.10} & \textbf{1.00} \\\hline
SVM &  94.47  & 91.90 & 8.10 & 96.80 & 3.20 & 0.923 \\\hline
NB & 88.01 & 85.15 & 14.85 & 90.55 & 9.45 & 0.939 \\\hline
\end{tabular} 
			}
\end{table}
\begin{table}[!t]
		\caption{Teacher and student models comparison using the RF classifier.}
		\label{tab:Performance-Evaluation-of}
		\centering
	\scalebox{0.9}{
			\begin{tabular}{l|l|l|l|l|l|l}
				
				\hline \hline
				 Model & ACC (\%)  & TPR(\%) & FPR (\%) & TNR (\%) & FNR (\%) & AUC \\\hline
			Teacher & 99.66 & 99.55 & 0.50 & 99.73 & 0.27 & 1.00 \\\hline
	      	Student & 99.59 & 99.45 & 0.55 & 99.71 & 0.28 & 0.99 \\\hline
			
			\end{tabular} 
			 }
	\end{table}

\section{Conclusions}
\label{Conclusion}

Intrusion detection applications are data hungry and training an effective model requires a huge amount of labeled data. In this paper, a knowledge transfer methodology for generating a shareable intrusion detection model has been presented to address the problem of enabling the research community to benefit from datasets owned by the intrusion detection agencies without directly sharing sensitive data. We believe that through mimic learning, a network detection student model can be trained and shared with outside communities to enable knowledge sharing with fewer privacy concerns. The performance evaluation of both the student model and the teacher model show nearly identical performance, which we consider to be an indication of the success of our mimic learning technique for transferring the knowledge from the teacher model to the student model using an unlabeled public data. 
\section*{Acknowledgment}  
This research work was financially supported in part by NSF CNS \#1619250. In addition, parts of this paper, specifically Sections I, II, IV, and V, were made possible by NPRP grants NPRP10-1223-160045 from the Qatar National Research Fund (a member of Qatar Foundation). The statements made herein are solely the responsibility of the authors.
   
\bibliographystyle{IEEEtran}
\bibliography{CC} 
    
\end{document}